# The evolution of cooperation and diversity by integrated indirect reciprocity

Tatsuya Sasakia[1]*, Satoshi Uchida[2], Isamu Okada[3], and Hitoshi Yamamoto[4]

[1] Department of Community Development, Koriyama Women's College, Koriyama-city, Fukushima, Japan

[2] Research Center for Ethi-Culture Studies, RINRI Institute, Tokyo, Japan

[3] Department of Business Administration, Soka University, Hachioji-city, Tokyo, Japan

[4] Faculty of Business Administration, Rissho University, Tokyo, Japan

* To whom correspondence should be addressed: t.sasaki@koriyama-kgc.ac.jp

## Abstract

Indirect reciprocity is one of the major mechanisms for the evolution of cooperation in human societies. There are two types of indirect reciprocity: upstream and downstream. Cooperation in downstream reciprocity follows the pattern, 'You helped someone, and I will help you'. The direction of cooperation is reversed in upstream reciprocity, which instead follows the pattern, 'You helped me, and I will help someone else'. In reality, these two types of indirect reciprocity often occur in combination. However, upstream and downstream reciprocity have mostly been studied theoretically in isolation. Here, we propose a new model that integrates both types. We apply the standard giving-game framework of indirect reciprocity and analyze the model by means of evolutionary game theory. We show that the model can result in the stable coexistence of altruistic reciprocators and free riders in well-mixed populations. We also found that considering inattention in the assessment rule can strengthen the stability of this mixed equilibrium, even resulting in a global attractor. Our results indicate that the cycles of forwarding help and rewarding help need to be established for creating and maintaining diversity and inclusion in a society.

# Introduction

Reciprocal cooperation is an indispensable part of sustainable societies. Even nearly half a century after Trivers' seminal work (1) on reciprocal altruism, the exploration of game-theoretical models for the evolution of cooperation through reciprocity remains at the forefront of evolutionary biology and the social sciences. Because helping is costly, self-interested individuals will free ride on others, so unconditional cooperation is unlikely to evolve. Therefore, the standard paradigm in the evolution of cooperation is a type of cooperation that is conditional on the degree of the other party's cooperativeness, as in reciprocal cooperation.

To succeed in competition with free riders, cooperative reciprocators require enough cognitive capacity to effectively process information for discriminating non-free riders from free riders. When the interaction consists of iterated rounds between the same pair of individuals, reciprocity is often in the form of direct reciprocity (1-3). Direct reciprocity is expressed as follows: A helps B, and then B helps A. Direct reciprocity requires memorising what the co-player and one's self did for each other in the past rounds of the iteration. In the absence of such iterations—as in the case of generalized exchange (4)—reciprocity should be indirect (5-9). Indirect reciprocity extends closed pairwise interactions to relationships involving external third parties. Implementing indirect reciprocity thus requires knowing about what players who may be involved did to others or had done to them by others in the past, such as by observing the co-player directly or using reputation systems.

There are two types of indirect reciprocity: upstream and downstream (7). Downstream reciprocity, on the one hand, can be expressed as follows: B helps C, and then A helps B (Fig. 1**b**). In other words, the response to B helping C was not C helping B directly but B being helped by a third party A, who observed B helping C and consequently evaluated B positively; this led to B being helped by A or another party who was influenced by B's positive evaluation—for instance through gossip or reputation (10-12). This is called 'rewarding reputation' (13). Therefore, downstream reciprocity uses reputation to identify partners with whom to cooperate. The motivation for such a reputational mechanism in downstream reciprocity, thus, is often described, as follows: 'If I help you, then I will be deemed good, and then someone will help me'. This is called 'reputational giving' (14).

Upstream reciprocity, which is expressed as A helps B, and then B helps C (Fig. 1**a**), is characterised by the logic of not choosing the partners with whom to cooperate. This differs from the logic behind downstream reciprocity, which is based on conditional cooperation. Upstream reciprocity is a chain of altruistic behaviors (15), called 'paying it forward' (16-18), that increases driving forces such as gratitude (13,18-22) or a sense of indebtedness (22), rather than the expectation of direct or indirect reward. In the eyes of a third party, however, emotional behavior can be viewed as a kind of reputational one, and vice versa. These motivations for reciprocity can easily be intertwined with each other when evaluated.

Although upstream and downstream reciprocity are commonly observed behaviors in experimental settings and field research (23-27), evolutionary game theory predicts that natural selection can favor downstream reciprocity but not upstream reciprocity, for which no supportive mechanism exists (7,28-30). Notably, also common is that different types of reciprocal mechanisms can be applied in tandem for promoting cooperation (14). Baker and Bulkley (13) suggested that rewarding reputation and paying it forward can reinforce each other as complementary mechanisms. A recent experimental study reported that in situations in which downstream reciprocators can provide help as a reward, those who pay it forward can become more likely to forward the help received (22). However, upstream and downstream reciprocity are theoretically studied mostly in isolation, and the impact of their interplay on the evolution of cooperation is still unknown. This is the riddle to be solved.

Here, we present a new model that integrates both types of indirect reciprocity. By using the model, we will show that stable coexistence between reciprocal altruists and free riders can be achieved by a method based only on indirect reciprocity without incorporating other mechanisms such as direct or spatial reciprocity (29). Specifically, we attempted to implement the virtuous cycle of paying it forward and rewarding reputation (13), as follows (see Fig. 1**c**). Let B be the modeled integrated reciprocator, who can act as either an upstream or downstream reciprocator. First, assume that D helps E; witnessing this, the integrated reciprocator B deems D good and rewards them by helping them as a downstream reciprocator. Furthermore, if A is another integrated reciprocator who already deemed B good, they will try to reward B by helping them as well. Then, B will forward the help received to someone else (C) as an upstream reciprocator. This, again, may lead to B being rewarded by another witnessing integrated reciprocator. After that, the reactive cycle of forwarding and rewarding among integrated reciprocators may continue in the same way.

It should be recalled that the chain of unconditional helping by upstream reciprocators is easily terminated when facing a free rider (29). The re-activation of helping requires waiting for the fortune that a new chain will come. In contrast to this, it is expected in the model that helping is more likely to revive because of the intervention of selective rewarding, as is depicted above. Considering the interplay between forwarding and rewarding, as such, would be suitable as a first step towards comprehensive study of the interplay of upstream and downstream reciprocity. In the next section, we will model the integrated reciprocator by incorporating these forwarding and rewarding behaviors into an action rule for individuals and then analyze the model by means of evolutionary game theory.

## Results

**The setup.** We build the model on the basis of the giving game in a well-mixed population. We assume that, given any interaction event, two players are randomly selected from the population and then interact with each other in only one round. Who plays the role of the donor or recipient is determined by a coin toss. To simplify the analysis, we assumed that in each round, a player acts as both donor and recipient (31). When acting as a donor, each player is offered an option to help (C) or not (D). Helping leads to benefits $b$ for the recipient and costs $c$ for the donor, with $b > c > 0$. Not helping has no effect on either the donor or the recipient. Thus, this yields an example of the well-known prisoner's dilemma game (2). We also consider the probability of failing to implement an intended action—whether or not to help—denoted by $\epsilon$ (32).

We then applied the standard framework to study the evolution of indirect reciprocity on the basis of the giving game (33-35). The player's strategy is described using an action and an assessment rule. The action rule prescribes whether a player helps or not. After every round, each player acting as the donor is assigned a binary image of 'good' (G) or 'bad' (B) by following the assessment rule. Note that the player's image when acting as the recipient is assumed to remain unchanged. In this study, we consider public assessment, under which a representative observer monitors each game, enforces the assessment rule for updating images, and broadcasts information about the population. We allow each player to know the co-player's information regarding actions and images perfectly.

**Integrated reciprocators stepping forward.** To study the interplay of upstream and downstream reciprocity, we establish the circulation of forwarded and rewarded help, as in Fig. 1**c**. In this study, we examine integrated reciprocators that help conditionally on the integrated action rule (Table 1**a**), as follows. Those who received help in the previous round will help a potential recipient, irrespective of the recipient's image, and those who did not receive help in the previous round will help a potential recipient only if the recipient's image is good. In what follows, we analyze a minimalistic setting in which each individual can choose one of three strategies: unconditional cooperator (X), unconditional defector (Y), and integrated reciprocator (Z). Unconditional cooperator and defector always intend to help and not to help, respectively. The three strategies' relative frequencies are denoted by $x$, $y$, and $z$, respectively, with $x + y + z = 1$. We assume that in the learning process, strategies that earn a higher payoff are more likely to be imitated in the population. We studied this simple process by means of replicator dynamics (36) (see Materials and Methods for details). In what follows, we will present the results of the basement model (Model I) and tuned one (Model II).

**Model I: Stable coexistence of the good and the bad.** We first developed Model I by considering the simplest assessment rule: those who help are deemed good, and those who do not are deemed bad (Table 1**b**). This is just the well-known *scoring* rule (37-39). As shown in Fig. 2, Model I can stabilize the intermediate level of cooperation in a mixed state of reciprocators and defectors (at P in Fig. 2**a,b**). In maintaining the coexistence, while unconditionally forwarding help by the upstream-reciprocation part in the action rule (the upper row of Table 1**a**) can be exploited by defectors, this is compensated for by conditional rewarding from its downstream-reciprocation part (the bottom row of Table 1**a**). In this way, the riddle of the evolution of upstream reciprocity is resolved within indirect reciprocity.

Figure 2 shows more details of the evolution of the three strategies. We can see that the phase portraits have a continuum of fixed points in the interior of the simplex $\Delta = \{(x, y, z): x + y + z = 1\}$. Fascinating is the dimorphic dynamics between integrated reciprocators and defectors, seen along the edge YZ given by $x = 0$. In the case without errors (Fig. 2**a**), edge YZ generally consists of segment RZ, a basin of attractor P and segment YR, a continuum of boundary fixed points. Attractor P: $z = z_0$ is given by

$$z_0 = \frac{b - 2c}{b - c}$$

(1)

The location of attractor P asymptotically comes close to node Y ($z = 1$) as the benefit-cost ratio, $b/c$, increases. At attractor P the population average of the probability to help takes $-z_0^2 + 3z_0 - 1$. We see that curve PQ, a continuum of interior fixed points connecting points P and Q, divides the simplex. Turning to other boundaries, the dynamics between integrated reciprocators and cooperators along edge XZ are neutral, and the dynamics between cooperators and defectors along edge XY are dominated by defectors. In the long run, therefore, considering random fluctuations can lead the population to come in the vicinity of node Y, the 100%-state of defectors. In the case with errors (Fig. 2**b**), an attractor P and also a repeller Q can appear along edge YZ. While continuums of boundary fixed points disappear, those of interior fixed points, PQ, remain. The dynamics between reciprocators and cooperators become dominated by the former. Beside these changes, the evolutionary fate of the population in the long run remains similar, even more definitely converging to the 100%-defector state (see Materials and Methods for details).

While Model I succeeds in inducing the attractor between reciprocators and defectors, the equilibrium induced is not asymptotically stable (36) against the invasion of cooperators. Therefore, regardless of the presence or absence of errors, considering the random perturbation, the population will leave the coexistence state in the long run. This is similar to the evolution of indirect reciprocity by *scoring* (32,40,41). The lack of stability of the coexistence state can be understood as follows. The definition of goodness in Model I is based only on whether to help or not, thus giving rise to the infamous problem of 'unjustified defection' (9,42,43), when reciprocators refuse to help those who are deemed bad. In this case, the image of reciprocators becomes bad, and the chance of being rewarded by other reciprocators decreases. When such a chain reaction of unjustified defection and image downgrading occurs, the advantage of being a reciprocator rather than a cooperator is lost.

**Model II: Robustness against the invasion of cooperators.** To strengthen the stability of the coexistence state, we propose Model II with a tuned assessment rule (Table 1**c**). Under the new rule, only those who implement upstream reciprocity deserve to be rewarded by those who follow the action rule. That is, Model II better captures the virtuous cycle of forwarding and rewarding cooperation. Indeed, when receiving help in the previous round, those who help are deemed good, and those who do not, bad, and when receiving no help in the previous round, the donor's image remains unchanged (denoted as K in Table 1**c**) whether they help or not in the current round. The new assessment rule is a sort of *staying* rule (44,45) and has been invented for rewarding to be focused on upstream reciprocation.

Model II can result in the coexistence of reciprocators and defectors, which does not allow cooperators to invade. In fact, in striking contrast to Model I, the dynamics for Model II have no interior equilibria, whether with or without errors. Fig. 3 shows that all the interior orbits converge to the boundary of the simplex, particularly edge YZ. This is from the fact that if the rate of the implementation error, $\epsilon$, is sufficiently small, integrated reciprocators are better off than cooperators (that is, $P_Z - P_X > 0$ holds). In the case without errors (Fig. 3**a**), along edge YZ exists a unique fixed point, P: $z = z_0$, with the same coordinates as in Model I, and node Y is a saddle. At the attractor P the cooperation rate (the probability to do C) over the population is given by $-z_0^3 + 2z_0^2$. The dynamics on the other edges, XZ and XY, remain unchanged, as in Model I. This follows that P is even the global attractor. Then, turning to the case with errors (Fig. 3**b**), edge YZ can exhibit an attractor P and a repeller Q; thus, the population dynamics can be bistable, evolving either to the mixed state P or the 100%-defector state Y (see Materials and Methods for details).

The stability of the attractor P against the invasion of cooperators can be understood as follows. Assuming that the integrated reciprocator received no help in the previous round, even if they interacted with a co-player with a bad image, the reciprocator's image would not change due to the *staying* element (K) of the assessment rule in Model II (Table 1**c**). Thus, the occurrence of unjustifiable defection is prevented. This means that a reciprocator with a good image can keep that image and thus continue to deserve to be rewarded by other reciprocators.

## Discussion

This study is a point of departure in an uncharted region in the field of the evolution of indirect reciprocity. Theories that explain the evolution of upstream reciprocity have so far used models that combine other mechanisms, such as direct or spatial reciprocity, while excluding downstream reciprocity. Our model shows that it is possible to establish a global attractor that can sustain a high level of upstream reciprocation, even without assuming errors, by integrating it with downstream reciprocation. Surprisingly, in the attractor, the integrated reciprocators considered can coexist with all-out defectors while deterring the intrusion of unconditional cooperators. Indeed, none of the previous models of indirect reciprocity resulted in the stable coexistence of altruistic reciprocators and free riders for the harsh, prisoner's dilemma game in well-mixed populations. Instead, finding an attractor between conditional and unconditional cooperators has been intensively studied (33,46-50).

The results can be compared with what happens in the evolution of four strategies: unconditional cooperator, unconditional defector, upstream reciprocator, and downstream reciprocator. Indeed, our study shows that the replicator dynamics for the four strategies can only result in the bistable fate of the population, as in the evolution of downstream reciprocity. The state space is divided into two distinct regions by a continuum of stable and unstable fixed points, given by $z = c/(1-\epsilon)b$ (with $c/(1-\epsilon)b < 1$), that is, the planar set (Fig. 4). Considering the random perturbation thus leads the population to end up with the 100%-defector state. This reveals that the simple extension of the strategy space to upstream reciprocators, as such, has no effect on improving the stability of cooperation (see Materials and Methods for details).

Here, let us discuss of the role of errors. It has been established that conditional strategies that attempt to establish cooperation can be eroded by unconditional cooperation strategies—ironically, once full cooperation is established. In a fully cooperative regime, conditional cooperators cannot be distinguished from unconditional ones and are thus seen as neutral mutants. Once unconditional cooperators spread to some extent, the invasion of all-out defectors will take place. Hence, most models of the evolution of conditional cooperation have considered errors that lead conditional cooperators to refuse to help unconditional cooperators. By this, conditional cooperators can be better off than unconditional cooperators. Considering errors has hitherto been essential in stabilising conditional cooperation (34,48,51-53). The need for errors to maintain cooperation has thus been a sort of necessary evil in the evolution of indirect reciprocity. In this regard, our results are not based on an artificial extrapolation of error factors. The significance of this study is that it demonstrates that the coexistence of altruists and free riders can be endogenously established through an evolutionary process. The asymptotic stability of this coexistence is a merit that stems from integrating upstream and downstream reciprocity and not by considering downstream reciprocity in isolation.

One important issue we have left out is that the stable coexistence established in Model II can become unstable against the invasion of 'pure' upstream reciprocity (Table 2**a**). Pure upstream reciprocators are those who can free ride on the costly rewarding by the integrated reciprocators. To deal with this issue, a considerable countermeasure would be updating the assessment rule so as to downgrade pure upstream reciprocators. We also remark on another type of free rider, those who only employ downstream reciprocity (Table 2**b**) and thus who can free ride on the costly unconditional forwarding of help. Our analysis suggests that the coexistence established can be stable against the invasion of pure downstream reciprocators (see Materials and Methods for details).

Other high demand issues to address include systematic exploration of integrated assessment rules, extension to negative reciprocity or paying forward greed (54,55), combination of upstream reciprocity with more complex downstream reciprocity, such as the leading eight norms (35,56,57), application of private assessment in norm ecosystem (58-64), or further integration with other types of reciprocity (29,65,66) or sanctioning systems (67,68)

In human societies, the coexistence of individuals with diverse degrees of cooperativeness is commonly observed, and maintaining inclusion and diversity is one of the key factors for sustainable development. While previous research on the evolution of cooperation by reciprocity has mostly focused on exclusively establishing the monomorphic state with full cooperation, little has been revealed about the conditions under which the polymorphic state with high and low cooperation can evolve (69-71). We believe that this study can pave the way for further research to facilitate and strengthen social diversity.

## Materials and Methods

**Evolutionary dynamics and image dynamics.** We will analyze the model by means of evolutionary game theory and investigate the replicator dynamics for a set of strategies considered. We thus assume an infinitely large population and its slow evolution, such that the composition of the population may be supposed to stay without changes in consecutive rounds. The replicator dynamics are given, in general, by $ds/dt = s(P_S - P)$, in which $s$ denotes the relative frequency of individuals who employ strategy $S$, $P_S$ the expected payoff per round for strategy $S$ ($P_S$ is determined after playing the infinitely large number of rounds), and $P$ the average payoff over the population, given by $\sum_S s P_S$.

As the first step, let us investigate the dynamics for the three strategies: unconditional cooperator (X), unconditional defector (Y), and integrated reciprocators (Z). We denote these relative frequencies by $x$, $y$, and $z$, respectively. Thus, $x + y + z = 1$ and $P = xP_X + yP_Y + zP_Z$. We also describe the relative frequency of those who have a good image within each strategy subpopulation by $g_S$ with $S \in \{X, Y, Z\}$. We denote the frequency of the good over the whole population by $g = xg_X + yg_Y + zg_Z$.

Here, we introduce a minimalistic framework that can deal with the interplay of upstream and downstream reciprocity, by using the following, called the generalized first-order action and assessment rule. The generalized first-order assessment rule is given in the following matrix:

$$\begin{array}{cc} \text{in/out} & \begin{array}{cc} \text{give C} & \text{give D} \end{array} \\ \begin{array}{c} \text{received C} \\ \text{received D} \end{array} & \begin{pmatrix} g(C, C) & g(C, D) \\ g(D, C) & g(D, D) \end{pmatrix} \end{array}, \qquad (2)$$

in which each element $g(a, b)$ denotes the probability that the focal player who received action $a$ in the previous round and then gives action $b$ in the current round, with $a, b \in \{C, D\}$, are deemed good. This matrix is a function of what the focal player does and what was done to the focal player and thus can cover the first-order assessment rule such as *scoring* (Table 1**b**)

In the equilibrium state (attained by starting from the state in which all have a good image $g_S$) the frequency of the good for each strategy should satisfy the following:

$$g_S = \sum_{a,b \in \{C,D\}} u_S(a) \, v_S(b) \, g(a, b)$$
$$= u_S(C)v_S(C)g(C, C) + u_S(C)v_S(D)g(C, D) + u_S(D)v_S(C)g(D, C) + u_S(D)v_S(D)g(D, D)' \qquad (3)$$

in which $u_S(i)$ and $v_S(i)$ denote the probabilities that the focal player with strategy $S$ receives action $i$ and that the focal player with strategy $S$ gives action $i$, respectively, in a given round. Thus, $u_S(D) = 1 - u_S(C)$ and $v_S(D) = 1 - v_S(C)$.

Then, we give the generalized first-order action rule by the following matrix:

$$\begin{array}{cc} & \begin{array}{cc} \text{Good} & \text{Bad} \end{array} \\ \begin{array}{c} \text{received C} \\ \text{received D} \end{array} & \begin{pmatrix} p_S(C, G) & p_S(C, B) \\ p_S(D, G) & p_S(D, B) \end{pmatrix} \end{array}, \qquad (4)$$

in which each element $p_S(a, i)$ denotes the probability that the focal player who received action $a \in \{C, D\}$ in the previous round and then is given an opponent with image $i \in \{G, B\}$ implements action C as a potential donor in the current round. This framework can cover the fundamental action rules: integrated reciprocity (Table 1**a**), upstream reciprocity (Table 2**a**), and downstream reciprocity (Table 2**b**).

By using the notations in Eqn (4), the probability that a donor with strategy $S$ implements C to (or helps) a recipient with strategy $T$ is given by

$$u(S,T) = \sum_{a\in\{C,D\}, i\in\{G,B\}} u_S(a)\, g_T(i)\, p_S(a,i)$$
$$= u_S(C)g_T(G)p_S(C,G) + u_S(C)g_T(B)p_S(C,B) + u_S(D)g_T(G)p_S(D,G) + u_S(D)g_T(B)p_S(D,B)$$
$$, (5)$$

in which $g_T(G) := g_T$ and thus $g_T(B) = 1 - g_T(G)$. This yields that

$$u_S(C) = \sum_{S'} s'\, u(S', S), \tag{6}$$

and

$$v_S(C) = \sum_{S'} s'\, u(S, S'). \tag{7}$$

Therefore, for the minimalistic setting with the strategy space {X, Y, Z}, we have

$$\begin{aligned}
u_X(C) &= x(1-\epsilon) + y\epsilon + z[u_Z(C)(1-\epsilon) + u_Z(D)(g_X(G)(1-\epsilon) + g_X(B)\epsilon)], \\
u_Y(C) &= x(1-\epsilon) + y\epsilon + z[u_Z(C)(1-\epsilon) + u_Z(D)(g_Y(G)(1-\epsilon) + g_Y(B)\epsilon)], \\
u_Z(C) &= x(1-\epsilon) + y\epsilon + z[u_Z(C)(1-\epsilon) + u_Z(D)(g_Z(G)(1-\epsilon) + g_Z(B)\epsilon)],
\end{aligned} \tag{8}$$

and

$$\begin{aligned}
v_X(C) &= 1 - \epsilon, \\
v_Y(C) &= \epsilon, \\
v_Z(C) &= u_Z(C)(1-\epsilon) + u_Z(D)[g(1-\epsilon) + (1-g)\epsilon].
\end{aligned} \tag{9}$$

By solving Eqs. (3,8,9), we can obtain $g_S(G)$, $u_S(C)$, and $v_S(C)$ for each point $(x, y, z)$ of the state space $\Delta$.

We assume that the image dynamics in Eqs. (3,5) are so fast that the replicator dynamics can be determined by the expected payoffs which depend on $u_S(C)$ and $v_S(C)$ in the equilibrium state of the image dynamics. We also assume that the image dynamics start from a situation in which all individuals have a good image. The expected payoffs for strategies are given by

$$P_S = bu_S(C) - cv_S(C), \tag{10}$$

**Model I.** From the assessment rule that those who help are deemed good (Table 1**b**), we have

$$g_S = v_S. \tag{11}$$

Thus, substituting Eq. (11) into Eq. (8) yields

$$\begin{aligned}
P_Z - P_Y &= (g_Z(G) - g_Y(G))[P_X - P_Y], \\
P_Z - P_X &= (g_Z(G) - g_X(G))[P_X - P_Y],
\end{aligned} \tag{12}$$

in which

$$P_X - P_Y = bz(1 - u_Z(C)) - c, \tag{13}$$

holds. The zero set of $P_X - P_Y$ as a function of $(x, y, z)$ provides a continuum of fixed points for the replicator dynamics in the interior of the two-dimensional state space $\Delta$. This is what the interior curve PQ describes in Figs. 2**a,b**. We first focus on the case without errors (Fig. 2**a**). From Eqs. (12,13), we have on edge YZ that for segment ZR with $(3-\sqrt{5})/2 < z \leq 1$, the fraction of the good converges to

$$g_Z(G) = -\frac{z^2 - 3z + 1}{z}, \tag{14}$$

or otherwise, for segment RY with $0 \leq z < (3 - \sqrt{5})/2$, to $g_Z = 0$. Hence, the fraction of the good over the whole population (that is, the frequency of those who cooperate) is $g = (1 - z_0)g_Y(G) + z_0 g_Z(G) = -z_0^2 + 3z_0 - 1$. Substituting Eq. (14) into Eq. (13) yields the zero set of Eq. (13) on segment ZR, which is given by $z_0 = (b - 2c)/(b - c)$ in Eqn (1). This yields that the point P with $z = z_0$ is an attractor with a basin, ZR. On the other side, segment RY consists exclusively of fixed points, along which $g_Z = g_Y = 0$ yield $P_Z = P_Y = 0$. Turning to the dynamics along edge XZ, we have $g_Z = g_X = 1$, and thus $P_Z = P_X$. Hence, it follows that the dynamics of reciprocators and cooperators are neutral. On edge XY, it is obvious that $z = 0$ yields $P_X - P_Y = -c < 0$ and thus that defectors dominate cooperators.

We then examine the case with errors (Fig. 2**b**). Using numerical simulations, we see that an attractor P and also a repeller Q can appear along edge YZ in general. Since the error rate is non-zero, the fraction of the good among reciprocators, $g_Z(G)$, can always take the non-zero value. Similarly, $g_Z(G)$, never takes its full value. As a result, no continuum of boundary fixed points appears along the boundary of the state space. In contrast to this, Eqs. (12,13) hold, irrespective of the presence or absence of errors, and thus a continuum of interior fixed points remains. When considering neutral drift or random perturbation, in particular in the case with errors, the population in the long run will converge to the 100%-defector state (node Y). Of interest is that the global dynamics for Model I have some similarity with those for *scoring* (31).

**Model II.** By the *staying* element in the assessment rule (Table 1**c**), in the equilibrium state of the image dynamics, we have the following equations:

$$\begin{aligned} g_X(G) &= u_X(C)(1 - \epsilon) + u_X(D)g_X(G), \\ g_Y(G) &= u_Y(C)\epsilon \quad\quad\quad + u_Y(D)g_Y(G), \\ g_Z(G) &= u_Z(C)(1 - \epsilon) + u_Z(D)g_Z(G), \end{aligned} \quad (15)$$

which obviously lead to the following constant values:

$$g_X(G) = 1 - \epsilon, g_Y(G) = \epsilon, \text{ and } g_Z(G) = 1 - \epsilon. \quad (16)$$

In striking contrast to Model I, the replicator dynamics for Model II have no interior equilibrium in the state space, and we thus can see that all interior orbits will converge to the boundary of the state space (Figs. 3**a,b**). Indeed, the payoff difference between reciprocators and cooperators is given by

$$P_Z - P_X = c(1 - u_Z(C))(1 - g_Z(G))(1 - 2\epsilon), \quad (17)$$

in which $(1 - u_Z(C))(1 - g_Z(G)) \neq 0$ holds in the interior state space, yielding $P_Z - P_X > 0$ for the sufficiently small errors with $\epsilon < 1/2$.

Next, let us check the dynamics between reciprocators and defectors along edge YZ. For $x = 0$, we have that

$$P_Z - P_Y = bz(1 - u_Z(C))(1 - 2\epsilon) - c[u_Z(C) + u_Z(D)g_Z(G)], \quad (18)$$

and, furthermore in the case without errors ($\epsilon = 0$),

$$P_Z - P_Y = -z^2(b - c) + z(b - 2c). \quad (19)$$

Thus, for $b > 2c$, point P with $z = z_0$ the same as in Eq. (1) becomes an attractor along edge YZ. We note also that the dynamics along edges XZ and XY remain unchanged from those for Model I. From these reasons, in the case without errors, it follows that all interior orbits will converge to P and thus that P is the global attractor (Fig. 3**a**). By using Eq. (9), we can have the probability that reciprocators give C, $v_Z(C)$ be equal to $z_0(2 - z_0)$, and thus its population average is $z_0^2(2 - z_0)$. In the case with errors, it turns out that an attractor P and a repeller Q appear simultaneously on edge YZ. Hence, the replicator dynamics have only two local attractors, P and node Y. As a result, the global dynamics are bistable: the population will converge to either P or Y (Fig. 3**b**).

**Stability of the attractor P against the invasion of pure downstream reciprocators in Model II.** Here, we prove in the case without errors that a rare mutant of pure downstream reciprocators (W) is worse off than the resident population consisting of defectors (Y) and reciprocators (Z). Consider pure downstream reciprocators (PDR) who employ the action rule by Extended Data Table 1**b** and the assessment rule by Table 1**c**. We first note that $g_W(G) = u_W(C)g + u_W(D)g_W(G)$, and thus $g_W(G) = z$ on edge YZ. Using this, we calculate the probability for PDR to receive C as $u_W(C) = (1-z) \cdot 0 + z(u_Z(C) + u_Z(D)g_W(G)) = z(u_Z(C) + u_Z(D)z)$, and the probability for PDR to give C as $v_W = (1-z) \cdot 0 + zg_Z = z$.

Similarly, the probability for integrated reciprocators to receive C is given by $u_Z = (1-z) \cdot 0 + z(u_Z(C) + u_Z(D)g_Z(G)) = z > u_W$, and the probability for integrated reciprocators to give C, $v_Z(C)$, is equal to $v_W(C)$. Therefore, it follows that the expected payoff for the mutant PDR, $P_W = bu_W(C) - cv_W(C)$, is smaller than that for the resident reciprocators, $P_Z = bu_Z(C) - cv_Z(C)$. That is, the mutant PDR is not selected for in the residents along edge YZ (including P).

**Cooperator, defector, upstream reciprocator, and downstream reciprocator.** We also explore the evolution of four strategies: unconditional cooperator, unconditional defector, upstream reciprocator, and downstream reciprocators. Downstream reciprocator intends to help a recipient, if the recipient helped someone else in the previous round. If the recipient did not help, downstream reciprocator intends not to help (Table 2**b**). Upstream reciprocator intends to help a recipient, irrespective of the recipient's image, if upstream reciprocator received help in the previous round. Otherwise, upstream reciprocator intends not to help (Table 2**a**).

We denote by $x$, $y$, $v$, and $w$ the relative frequencies of unconditional cooperator (X), unconditional defector (Y), upstream reciprocator (V), and downstream reciprocator (W), respectively. Thus, $x + y + v + w = 1$ and $P = xP_X + yP_Y + vP_V + wP_W$. The frequency of the good over the whole population is given by $g = xg_X + yg_Y + vg_V + wg_W$. Then, as in Eq. (8), we can have the following equations to define $u_S(C)$ recursively:

$$\begin{aligned}
u_X(C) &= x(1-\epsilon) + y\epsilon + v[u_V(C)(1-\epsilon) + u_V(D)\epsilon] + w[g_X(G)(1-\epsilon) + g_X(B)\epsilon], \\
u_Y(C) &= x(1-\epsilon) + y\epsilon + v[u_V(C)(1-\epsilon) + u_V(D)\epsilon] + w[g_Y(G)(1-\epsilon) + g_Y(B)\epsilon], \\
u_V(C) &= x(1-\epsilon) + y\epsilon + v[u_V(C)(1-\epsilon) + u_V(D)\epsilon] + w[g_V(G)(1-\epsilon) + g_V(B)\epsilon], \\
u_W(C) &= x(1-\epsilon) + y\epsilon + v[u_V(C)(1-\epsilon) + u_V(D)\epsilon] + w[g_W(G)(1-\epsilon) + g_W(B)\epsilon].
\end{aligned} \quad (20)$$

By solving Eqs. (3,9,20), we can have $g_S(G)$, $u_S(C)$, and $v_S(C)$. Substituting these into Eq. (10) allows us to calculate the payoffs and thus the replicator dynamics. For Model II, $v_S$, the probability that a player with strategy $S$ gives C is given by:

$$\begin{aligned}
v_X(C) &= 1 - \epsilon, \\
v_Y(C) &= \epsilon, \\
v_V(C) &= u_V(C)(1-\epsilon) + u_V(D)\epsilon, \\
v_W(C) &= g(1-\epsilon) + (1-g)\epsilon.
\end{aligned} \quad (21)$$

Fig. 4 describes the evolution of the four strategies by the replicator dynamics. Fig. 4**a** shows the boundary dynamics on each face. On the X-Y-V plane, defector is dominant. For other three faces (X-Y-W, X-W-V, and Y-W-V), there can exist a continuum of interior fixed points if $c/(1-\epsilon)b < 1$. We see also that the edge dynamics between downstream and upstream are neutral. Therefore, the random shock can bring the population eventually to node Y, which is the homogeneous state for defectors.

Fig. 4**b** shows the interior dynamics. If $c/(1-\epsilon)b < 1$, there exists an intersection of the plane and the 3D simplex $\Delta_4 = \{(x, y, v, w): x + y + v + w = 1\}$. Otherwise, there is no interior fixed point in $\Delta_4$. Fig. 4**b** shows the intersection consists of stable and unstable fixed points. Depending on the initial conditions, the population may first evolve to a stable point within the planar continuum of fixed points. Whatever the initial conditions, the random perturbation can still lead the population to finally converge to node Y.

## Acknowledgments

This work was supported by JSPS KAKENHI Grant Numbers JP19H02376 (I.O, H.Y), JP20K20651 (I.O), JP21H01568 (I.O, H.Y), JP21KK0027 (I.O, H.Y), JP22H03906 (H.Y, I.O), and JP19K21570 (H.Y). The funders have/had no role in study design, data collection and analysis, decision to publish or preparation of the manuscript. We are grateful to Å. Brännström, U. Dieckmann, Y. Nakai, H. Ohtsuki, and N. Takahashi for their comments.

# Figures and Tables

## Figure 1

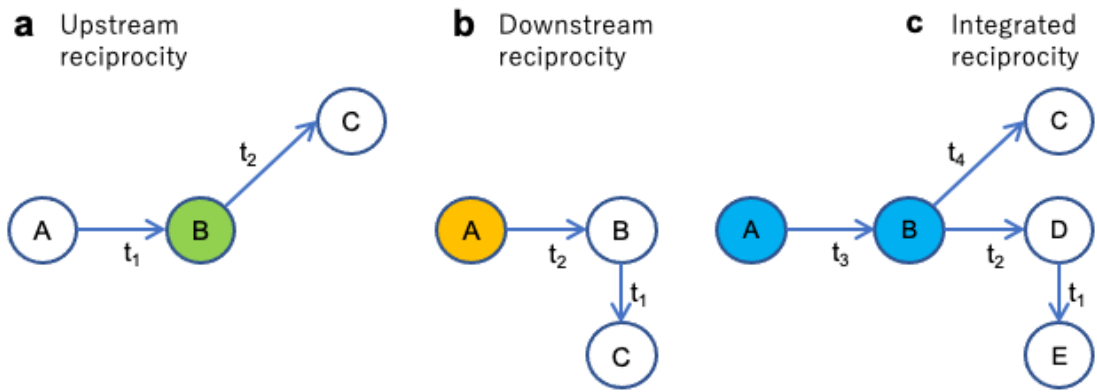

**Figure 1. Three types of indirect reciprocity. a,** Upstream reciprocity: first A helps B, and then B (upstream reciprocator) forwards the help received to B. **b,** Downstream reciprocity: first B helps C, and then A (downstream reciprocator) rewards B by helping. **c,** Integrated reciprocity: first D helps E, then B (integrated reciprocator) rewards D by helping; next, given that the other integrated reciprocator, A, rewards B by helping, B, moved by this, will forward the help received to C. Another integrated reciprocator may also subsequently reward B, who is moved by this and again forwards the help received to someone else.

**Figure 2**

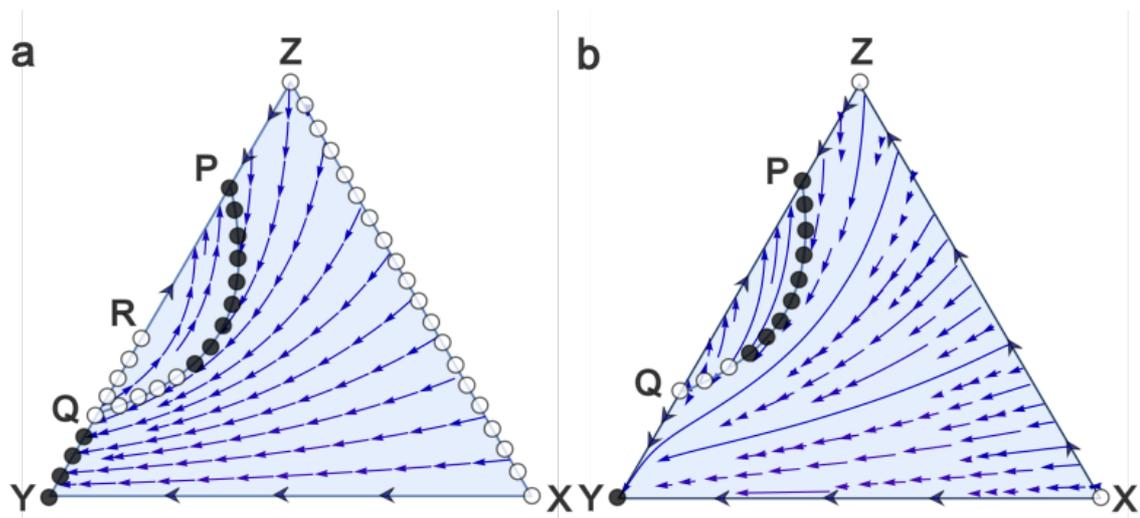

**Figure. 2. Evolution of integrated reciprocity for Model I.** Panels **a** and **b** depict phase portraits of the replicator dynamics for unconditional cooperator X, unconditional defector Y, and integrated reciprocator Z, without and with errors, respectively. The triangles describe a simplex of the state space $\Delta = \{(x, y, z): x + y + z = 1\}$. Each node (X, Y, or Z) of the triangle corresponds to the homogeneous state of each strategy ($x$, $y$, or $z = 1$, respectively). Moreover, filled and empty circles denote stable and unstable fixed points, respectively. **a,** Without errors, the simplex has three continuums of fixed points: XZ, RY, and PQ, among which only PQ remains, even when assuming errors in **b**. Whether errors are present or absent, considering random shock on the population composition, the presence of a continuum of interior fixed points, PQ, prevents the population from staying at the boundary attractor P. **b,** In particular, the population will eventually converge to node Y (100%-defector state). Parameters: $c = 1$, $b = 5$, (**a**) $\epsilon = 0$, and (**b**) $\epsilon = 0.05$.

**Figure 3**

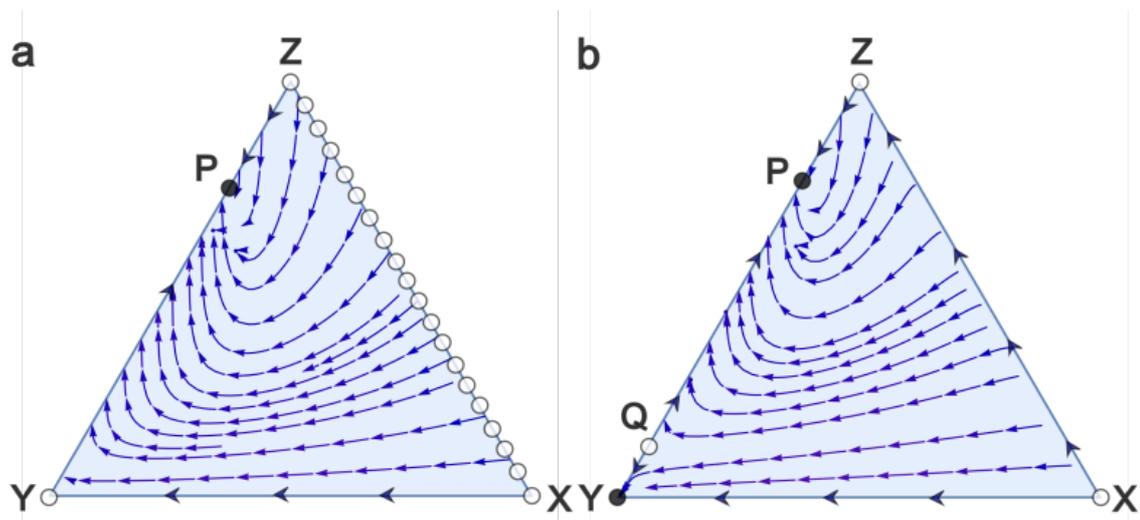

**Figure 3. Evolution of integrated reciprocity for Model II.** Panels **a** and **b** depict phase portraits of the replicator dynamics for unconditional cooperator X, unconditional defector Y, and integrated reciprocator Z, as in Fig. 2. **a,** Without errors, the dynamics show the global attractor P along edge YZ. At P, integrated reciprocators and defectors stably coexist. **b,** The global dynamics become bistable: the population will eventually converge to either the local attractors P or Y (100%-defector state). Parameters: $c = 1$, $b = 5$, **(a)** $\epsilon = 0$, and **(b)** $\epsilon = 0.1$.

**Figure 4**

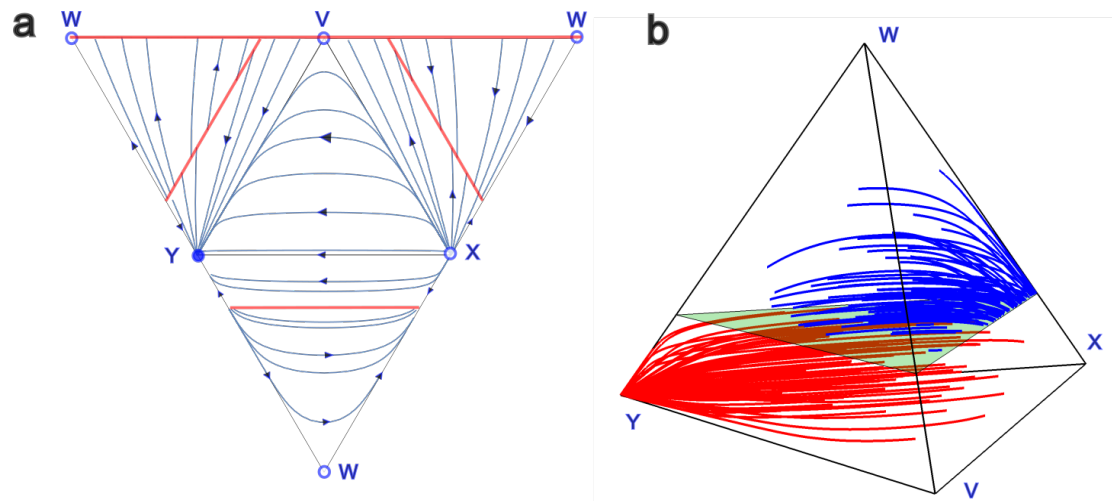

**Figure 4. Evolution of traditional four strategies.** Panels **a** and **b** depict phase portraits of the replicator dynamics for unconditional cooperator X, unconditional defector Y, upstream reciprocator V, and downstream reciprocator W, on the surfaces and in the interior space, respectively. The tetrahedral simplex in **b** denotes the state space $\Delta_4 = \{(x, y, v, w): x + y + v + w = 1\}$. Each node (X, Y, V, or W) corresponds to the homogeneous state for each strategy. The tetrahedral simplex intersects a planar set that consists of stable and unstable fixed points. Red solid lines in **a** denote a set of continuums of boundary fixed points, among which PQ, QR, and RP are the intersections between each triangular surface and the planar set. In **b,** the global dynamics show that, on the one hand, blue interior orbits are converging to points on the plane, and also that on the other hand, red interior orbits are attaining node Y, which indicates the local stability of node Y. Because of the planar set, while considering the random fluctuation, the evolution can end up with the 100%-defector state at node Y. Parameters: $c = 1$, $b = 5$, and $\epsilon = 0.1$.

# Table 1

**a.** Action rule

|  |  | Image of recipient | |
|---|---|---|---|
|  |  | G | B |
| **In the previous round** | received C | C | C |
|  | received D | C | D |

**b.** Assessment rule for Model I

|  |  | In the current round | |
|---|---|---|---|
|  |  | give C | give D |
| **In the previous round** | received C | G | B |
|  | received D | G | B |

**c.** Assessment rule for Model II

|  |  | In the current round | |
|---|---|---|---|
|  |  | give C | give D |
| **In the previous round** | received C | G | B |
|  | received D | K | K |

**Table 1. Action and assessment rules for integrated reciprocity.**

# Table 2

**a.** Action rule for upstream reciprocity

|  |  | Image of recipient | |
|---|---|---|---|
|  |  | G | B |
| **In the previous round** | received C | C | C |
|  | received D | D | D |

**b.** Action rule for downstream reciprocity

|  |  | Image of recipient | |
|---|---|---|---|
|  |  | G | B |
| **In the previous round** | received C | C | D |
|  | received D | C | D |

**Data Table 2. Action rules for upstream and downstream reciprocity.**